\documentclass{PoS}

\title{Generalized Parton Distributions: \\
	the experimental status}

\ShortTitle{Generalized Parton Distributions: the experimental status}

\author{\speaker{F.~Sabati\'e}, H.~Moutarde\\
        CEA Saclay - Irfu/SPhN\\
        E-mail: \email{franck.sabatie@cea.fr, herve.moutarde@cea.fr}}

\abstract{We review the experimental as well as the phenomenology status of Generalized Parton Distributions (GPDs), focusing on recent data on Deeply Virtual Compton Scattering and Deep Virtual Meson Production. We also describe the various strategies for extraction of GPDs from these data. Finally we conclude with a summary of we know about GPDs and the experimental outlook in the years to come.}

\FullConference{Sixth International Conference on Quarks and Nuclear Physics,\\
		April 16-20, 2012\\
		Ecole Polytechnique, Palaiseau, Paris}

\begin{document}

\section{Introduction}

Generalized Parton Distributions (GPDs) are physical observables which can provide deep insight about the internal structure of the nucleon and more generally, non-perturbative QCD. They contain the usual parton distribution functions (PDFs) and elastic form factors (FFs) as limiting cases or sum rules. In addition, GPDs allow to probe the nucleon as a 3-dimensional object, accessing for instance the quark orbital momentum and picturing the nucleon in quantum phase space.

Generalized Parton Distributions can be accessed through deep exclusive processes such as Deeply Virtual Compton Scattering (DVCS) or Deep Virtual Meson Production (DVMP). About fifteen years ago, Mueller, Ji, Radyushkin and others~\cite{Mueller:1998fv,Ji:1996nm,Ji:1996ek,Ji:1997gm,Radyushkin:1996nd,Radyushkin:1997ki}
showed that the DVCS reaction $\gamma^* p \to \gamma p$ can,
in the Bjorken limit, be factorized into a hard scattering kernel and a non-perturbative part, containing
information about the electromagnetic structure of the nucleon in terms of four twist-2 chiral-even GPDs $H$, $E$, $\widetilde{H}$ and $\widetilde{E}$.

These GPDs depend on 4 variables $(x,\xi,t;Q^2)$. $x$ characterizes the average light-cone momentum fraction of the
struck quark in the intermediate state (not directly accessible experimentally). $\xi$ is the longitudinal momentum fraction of
the transfer to the proton $\Delta=p-p'$ (where $p$ and $p'$ are the initial and recoil proton 4-vectors).
Finally, $t=\Delta^2$ is the standard Mandelstam variable representing the momentum transfer between the virtual
and real photons (or between the target and the recoil proton). The scale evolution of the GPDs ($Q^2$-dependence) has been worked out to next-to-leading order of $\alpha_S$ and beyond~\cite{Mueller:2005nz,Kumericki:2006xx}.

In the following, we will introduce the simple case of DVCS to show how this reaction gives access to GPDs. We will then describe a few recent results on DVCS and DVMP from the Jefferson Lab Hall A, CLAS, HERMES and COMPASS collaboration. We will describe the current efforts on GPD extraction and finally conclude on what we have learned so far, as well as what can be expected experimentally in the future.

\section{Deeply Virtual Compton Scattering, a tool to access GPDs}

The photon electroproduction $e p \to e p \gamma$ can occur either by radiation along
one of the electron lines (Bethe-Heitler or BH) or by emission of a real
photon by the nucleon (DVCS). The total amplitude $\mathcal T_{e p \gamma}$ is therefore the superposition of the BH and DVCS amplitudes:
\begin{equation}
\left | \mathcal T_{e p \gamma} \right |^2 =  \left | \mathcal T_{\rm BH} \right |^2+\left |
\mathcal T_{\rm DVCS} \right |^2+ \mathcal I \,,
\end{equation}

\noindent where $\mathcal T_{\rm DVCS}$ and $\mathcal T_{\rm BH}$ are the amplitudes
for the DVCS and Bethe-Heitler processes, and $\mathcal I$ denotes the
interference between these processes.

Using either a polarized beam or a longitudinally polarized target, two separate quantities can be extracted: the difference of cross section
for opposite beam helicities or opposite target spin and the total cross section, which at leading twist can be written respectively as:
\begin{eqnarray}
d\sigma^\rightarrow-d\sigma^\leftarrow & = & 2 \cdot\mathcal T_{BH} \cdot {\mathrm{Im}}(\mathcal T_{DVCS})\, ,\\
d\sigma^\rightarrow+d\sigma^\leftarrow & = & \left| \mathcal T_{BH}\right|^2 + 2 \cdot\mathcal T_{BH} \cdot {\mathrm{Re}}(\mathcal T_{DVCS}) + \left|T_{DVCS}\right|^2 \, ,
\end{eqnarray}
\noindent where the arrows correspond to the beam helicity. At low beam energy, the pure DVCS contribution is expected to be small with respect to the interference terms, which themselves are in general significantly smaller than the BH term. Note that the DVCS contribution to the difference of cross section only appears at higher twist. It is actually natural to express the DVCS amplitude $\mathcal T_{\rm DVCS}$ at leading twist more generally in terms of so-called Compton Form Factors (CFFs) which can be written at leading order as a function of the GPDs :
\begin{eqnarray}
\mathcal{F} 
& = & \int_{-1}^{+1} dx \, F(x,\xi,t) \left( \frac{1}{\xi-x-i\epsilon} - \frac{1}{\xi+x-i\epsilon} \right) \qquad (F = H \textrm{ or } E) \\
& = & \mathcal{P} \int_{-1}^{+1} dx \, F(x,\xi,t) \left( \frac{1}{\xi-x} - \frac{1}{\xi+x} \right) + i \pi \Big( F(\xi,\xi,t) - F(-\xi,\xi,t) \Big) \end{eqnarray}

The symbol $\mathcal{P}$ stands for Cauchy principal value. A CFF is complex-valued and we note $\mathop{\mathrm{Re}}$~CFF and $\mathop{\mathrm{Im}}$~CFF its real and imaginary parts. They are related by fixed-$t$ dispersion relations \cite{Teryaev:2005uj, Anikin:2007tx, Anikin:2007yh, Diehl:2007jb}, for example~:
\begin{equation}
\mathop{\mathrm{Re}}\mathcal{H}( \xi, t ) = 2 \mathcal{P} \int_0^{1} \frac{d\xi'}{\xi'} \frac{\mathop{\mathrm{Im}}\mathcal{H}( \xi', t )}{\frac{\xi^2}{\xi'^2}-1} + D(t)
\label{eq-dispersion-relation}
\end{equation}

\noindent where $D(t)$ is a subtraction constant related to the $D$-term \cite{Polyakov:1999gs}. However the $D$-term is poorly known and most of DVCS measurements are made in the region $\xi' \leq 0.5$. Using such dispersion relations thus relies on models and may introduce biases in the extraction of GPDs from DVCS data. For that reason the real and imaginary parts of CFFs are taken as independent in some fitting procedures although they should obey the equality (\ref{eq-dispersion-relation}) from first principles.

\section{Recent experimental results}

\begin{figure}
\begin{center}
\includegraphics[width=1\textwidth]{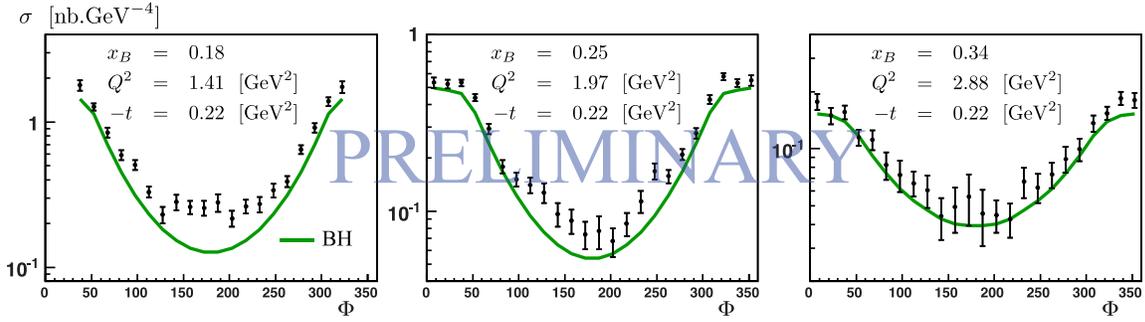}
\caption{Sample of preliminary CLAS data on photon electroproduction cross sections as a function of $\phi$ for three kinematical bins.}
\label{fig:DVCS-CLAS}
\end{center}
\end{figure}

Most of the recent experimental results come from DVCS and we will start with that process. The CLAS collaboration is at the final stage of releasing cross sections for photon electroproduction at 6~GeV in a wide kinematical range $0.2<x_B<0.6$, $1<Q^2<5$~GeV$^2$. Along with the older Hall~A cross section results \cite{MunozCamacho:2006hx}, these data will have a significant impact on the GPD extractions and will allow for better constraints on Im$\mathcal H$ and to a lesser extent Re$\mathcal H$. A sample of these preliminary data are shown on Figure~\ref{fig:DVCS-CLAS}. More information may be found in proceedings from this conference \cite{Proceed-HS}.

The CLAS collaboration is also analyzing data on longitudinal Target Spin Asymmetry for DVCS. These data are essential to better constrain the GPD $\widetilde H$ through the imaginary part of the corresponding CFF. More information may be found in proceedings from this conference \cite{Proceed-Silvia}.

The Jefferson Lab Hall A collaboration has taken DVCS data recently on both proton and deuteron. The main goal of this experiment is to separate the DVCS$^2$ term from the interference and $BH$ terms in the cross section using a beam energy scan. At HERMES energies, the DVCS$^2$ term was evaluated to be small, but at higher-$x_B$ and somewhat smaller $Q^2$, the conclusion may prove different. More information may be found in these proceedings \cite{Proceed-Roche}.

\begin{figure}
\begin{center}
\includegraphics[width=.8\textwidth]{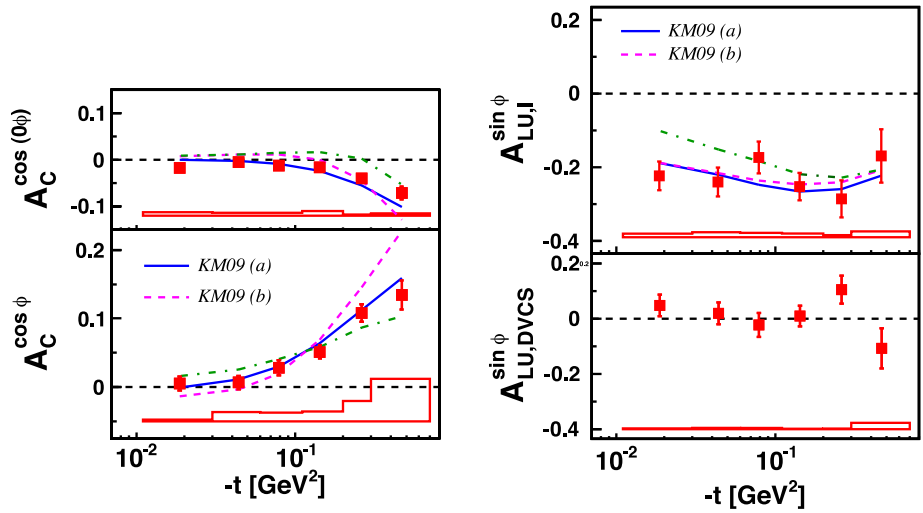}
\caption{Sample of newly published HERMES data on beam charge and spin asymmetry moments as a function of $-t$, taken from \cite{Airapetian:2012mq}.}
\label{fig:HERMES}
\end{center}
\end{figure}

The HERMES collaboration recently published updated results on beam charge and beam spin asymmetries for DVCS including newly analyzed 2005-2007 data \cite{Airapetian:2012mq}. The data from the 1996-2005 set are compatible with the 2005-2007 sample, and provide about twice the amount of data. A sample of the most significant moments of the beam charge and beam spin asymmetries published by the HERMES collaboration are shown on Figure~\ref{fig:HERMES}. Note that since the time of this conference, the HERMES collaboration released their data on kinematically complete events using their recoil detector. It was interesting to notice a large increase of the beam spin asymmetry in that case, due to the better handling of the dilution coming from inelastic events \cite{Airapetian:2012pg}.

Concerning meson electroproduction, two recent results are linked to GPD physics: COMPASS finalized its analysis of the $A_{\rm UT}$ asymmetry for the $\rho^0$. More information may be found in these proceedings \cite{Proceed-Schmidt}. CLAS also finalized its analysis of unseparated $\pi^0$ electroproduction cross sections \cite{Bedlinskiy:2012be}. They manage to prove that most of the cross section is transverse and that a GPD-based description requires the use of chiral-odd GPDs (also called transversity GPDs). Again, more information may be found in these proceedings \cite{Proceed-Kroll}.

\section{Considerations on GPD extraction}

Extractions of GPDs is a much more demanding task than the extraction of Parton Distribution Functions (PDF) or Form Factors (FF) due to the complex functional structures of GPDs. Moreover, we need to extract four functions $H$, $E$, $\widetilde{H}$ and $\widetilde{E}$ of three variables $(x, \xi, t)$ for each quark flavour ($u$, $d$ and $s$). The $Q^2$-dependence is governed by the QCD evolution equations. Building a flexible yet robust GPD parametrization is a challenging project and the problem is still open today. However, several groups have made attempts to fit GPDs (or CFFs) to data during the last few years. So far, all methods have in common to assume, the dominance of the DVCS twist-2 contribution and are designed to fit several physical observables at once. We will describe the different fitting methods in the following sections.

\subsection{Local fits of CFFs}

The first approach, pioneered in \cite{Guidal:2008ie} and used in \cite{Moutarde:2009fg,Guidal:2009aa,Guidal:2010ig,Guidal:2010de} assumes the independence of the real and imaginary parts of CFFs. Each kinematic bin $( x_B, t, Q^2 )$ is taken independently of the others, and seven values $\mathop{\mathrm{Re}}\mathcal{H}$, $\mathop{\mathrm{Im}}\mathcal{H}$, $\mathop{\mathrm{Re}}\mathcal{E}$, $\mathop{\mathrm{Im}}\mathcal{E}$, $\mathop{\mathrm{Re}}\widetilde{\mathcal{H}}$, $\mathop{\mathrm{Im}}\widetilde{\mathcal{H}}$ and $\mathop{\mathrm{Re}}\widetilde{\mathcal{E}}$ are extracted simultaneously. This method has the clear advantage of being almost model-independent. The drawback is that it does not give information on the extrapolation of the extracted CFF outside the data region.  In the following we will refer to these fits as \emph{local fits}.

\subsection{Global fits of GPDs}

In the spirit of the work done on PDFs and FFs, \emph{global fits} require a physically motivated parametrization of GPDs and deal with all kinematic bins at once. The main advantage is obvious: the ability to extrapolate outside of the data region, and therefore evaluate for instance Ji's sum rule ($t\to 0$) or more generally, study the 3D partonic structure of the nucleon ($\xi\to 0$). The free coefficients entering the expressions for GPDs are determined either from PDFs and FFs or from DVCS data. Two such studies have been reported recently for DVCS \cite{Kumericki:2009uq,Goldstein:2010gu}. Note that fixed-$t$ dispersion relations are used as a key ingredient in \cite{Kumericki:2009uq}.

\subsection{Hybrid fits of GPDs}

The \emph{hybrid} fitting procedure used in \cite{Moutarde:2009fg} is a combination of the previous two methods and has been applied with the main assumption of $H$-dominance and twist-2 accuracy. It involves a parametrization which fulfills the polynomiality condition of GPDs and includes $Q^2$ evolution at leading order in $\alpha_S$. Since this function is otherwise arbitrary, its specific form is \textit{a posteriori} validated by the quality of the fit. It makes it hazardous to extrapolate the extracted GPD outside the fitting domain as unphysical oscillations may occur. The model dependence is tested by a systematic comparison to local fits and an estimate of the systematic error induced by the $H$-dominance hypothesis. The good agreement of the local fits with respect to the global fits is a strong consistency check of this approach.

\subsection{Neural network fits of GPDs}

\emph{Neural network} fits had been successfully performed for PDFs but their use for GPD extraction is quite recent. First results are described in \cite{Kumericki:2011rz} within the $H$-dominance assumption. Although it is too early to assess the advantages and shortcomings of this approach, it is worth noting that it is a new development in the field of GPD extraction.

\section{Outlook and conclusion}

A sizeable data set is now available from H1, ZEUS, HERMES, Jefferson Lab CLAS, Hall A and COMPASS. The first step of GPD extraction and comparison to models \cite{KMS12} allowed to draw a few conclusions:
\begin{itemize}
\item $H$ dominance is not as clear as usually advertised for DVCS in the mid to high-$x_B$ region.
\item There is a necessity to go beyond simple leading-twist and leading-order formalism.
\item That being said, we have a reasonable idea of the size of $H$ for gluons, sea and valence quarks.
\item We have a rough idea of the size of $\widetilde H$ and $E$ for valence quarks
\item We have some limited clues on the size of $\widetilde H$ and $E$ for sea quarks.
\item We know almost nothing on $\widetilde E$ and the chiral-odd GPDs, but there is some progress.
\end{itemize}

The next stage is obvious : going from a “rough” to a “good” knowledge of the GPDs and finally get some insights about the structure of the nucleon. Clearly, accurate data on polarized cross sections in a large kinematical domain are absolutely crucial for that. Advances in theory and phenomenology are also needed, even though projects in GPD extraction have emerged recently with the wealth of new data. In the near and longer term future, we expect even more data to come:
\begin{itemize}
\item Short-term (2012-2013) : The CLAS collaboration still has a lot of potential with the cross section data as well as target spin asymmetries soon to be released. The Hall A collaboration is currently analyzing separated DVCS and $\pi^0$ data.

\item Mid-term (2014-2020+) : The COMPASS-II proposal was approved and a short DVCS run is planned for 2012, with a long run as soon as 2015. The CLAS12  and Hall A collaborations of Jefferson Lab have a huge DVCS and DVMP program including target and beam polarization observables. Note that an exciting new program on Timelike Compton Scattering was just accepted by the Jefferson Lab PAC \cite{Proceed-Jakub}.

\item Long-term (2025+) : An Electron Ion collider would be the ultimate tool for 3D nucleon imaging with a kinematical reach at high luminosity which would be unparalleled \cite{Proceed-Ent}.
\end{itemize}

\end{document}